# *The aqueous Triton X-100 - Dodecyltrimethylammonium bromide micellar mixed system. Experimental results and thermodynamic analysis.*

Patricio Serafini[1], Marcos Fernández-Leyes[1], Jhon Sánchez M.[1], Romina B. Pereyra[3], Erica P. Schulz[3], Gillermo A. Durand[4], Pablo C. Schulz[3] and Hernán A. Ritacco[(*)1,2].

[1] Instituto de Física del Sur (IFISUR-CONICET), Av. Alem 1253, Bahía Blanca (8000), Argentina

[2] Departamento de Física de la Universidad Nacional del Sur, Av. Alem 1253, Bahía Blanca (8000), Argentina.

[3] Instituto de Química del Sur (INQUISUR-CONICET) and Departamento de Química, Universidad Nacional del Sur.

[4] Planta Piloto de Ingeniería Química (PLAPIQI- CONICET-Universidad Nacional del Sur).

(*) Corresponding author: hernan.ritacco@uns.edu.ar

**ABSTRACT.**

The micellization process of the aqueous mixed system Triton X-100 (TX100) – Dodecyltrimethylammonium Bromide (DTAB) was studied with a battery of procedures: surface tension, static and dynamic light scattering and ion-selective electrodes. Results were also analysed with two thermodynamic procedures. The system shows some changes in its behaviour with changing the mole fraction of DTAB, $\alpha_{DTAB}$, in the whole surfactant mixture. For $\alpha_{DTAB} \leq 0.40$ micelles are predominantly TX100 with scarce solubilized DTA$^+$ ions, and TX100 acts as a nearly ideal solvent. In the range $0.50 \leq \alpha_{DTAB} \leq 0.75$ it seems that none of the components acts as a solvent, and above $\alpha_{DTAB} \approx 0.75$ there are abrupt changes in the size and electrophoretic mobility of micelles. These phenomena have been interpreted in the light of the thermodynamic results and some TX100-ionic surfactant mixtures of literature.

Keywords: Triton X-100 – Dodecyltrimethylammonium Bromide – Mixed Micelles – Dynamic Light Scattering – Static Light Scattering – Electrophoretic mobility – Equation Oriented Mixed Micelle Model (EOMMM)

**INTRODUCTION.**

Molecules that have the property of adsorbing spontaneously onto the interface between two immiscible fluid phases are known as surfactants[1,2] (surface active). Surfactants are small molecules that possess two distinct parts on their chemical structure: one with affinity to polar solvents (like water), the polar head, and the other with affinity to non-polar fluids, the hydrophobic tail. At a certain well defined concentration, surfactants self-assemble in bulk to form aggregates called micelles. The concentration at which this happens is called Critical Micelle Concentration (CMC), which is the most important characteristic of surfactants. The surfactant micellization is a cooperative process producing abrupt changes in many physical properties such as surface tension, light scattering, conductivity, etc. The measurement of those properties as a function of surfactant concentration permit the determination of the CMC [1,2].

Surfactant solutions are used in many technical applications such as enhanced oil recovery, detergency, pharmaceuticals, food, cosmetics, flotation mineral recovery, and pesticides, among others[1,2]. In almost all these applications surfactant mixtures are commonly used instead of pure surfactants, because these mixtures often have better performance, e.g lower CMC's, than one-component systems [1,3,4]. It is worth mentioning here that the mixtures include not only mixtures of different surfactants but also surfactants with polymers, polyelectrolytes, proteins, micro and nano particles [5,6]. These complex surfactant mixtures allow to formulate systems with designed properties, for instance, mixtures of a cationic surfactant with a thermorsponsive anionic polyeletrolyte was used for stabilizing liquid foams whose stability responds to an external stimuli [7]. In the case of detergent formulations, mixtures of ionic and nonionic surfactants are frequently used. Since the 1960's [8,9], it was demonstrated that mixtures of ionic surfactants with nonionic ones have improved features in regards of their applications [3,8–13]. The inclusion of these nonionic surfactants aid, for example, to reduce some undesirable interactions between ionic surfactants and the substrate, such as precipitation with polyvalent cations (mainly $Ca^{+2}$ or $Mg^{+2}$) or the tendency of electrostatic adsorption of cationic surfactants to natural negatively charged surfaces. In some applications such as germicide formulations, cationic surfactants are added to nonionic surfactants due to their biocide properties[14]. Although the synergistic effects when different surfactants are mixed have received great attention in view of its exploitation in designing mixtures with particular desirable properties, the basic processes are still relatively poorly understood at a detailed molecular level[14].

In surfactant mixtures, the determination of the composition of mixed micelles is a major problem since its value is fixed by the partition equilibria of the species between the aggregates and the surrounding medium. Because the mixed micelles composition is quite difficult to assess experimentally in a direct manner, it has to be estimated on the basis of a given thermodynamic model parameterized with physicochemical properties, mainly the critical micelles concentration (CMC) [1]. Among the mixtures composed by ionic and non-ionic surfactants, Carnero Ruiz and Aguiar [15] have studied three mixed surfactant systems, TX100 with Hexadecyltrimethylamonium Bromide (CTAB), Tetradecyltrimethylammonium Bromide (TTAB) and Dodecyltrimethylammonium Bromide (DTAB). At difference of the TX100-CTAB and the TX100-TTAB mixed systems, the TX100-DTAB system CMC values were not well modelled by the Regular Solution Theory [16] (RST or Rubingh' procedure). This fact and some other characteristics of this system which were different from those found with the other homologues studied in the above referred work makes the TX100-DTAB mixed system interesting to a detailed research.

In the present work we have studied a binary surfactant mixture formed by Triton-X100 (TX100), a non-ionic surfactant, and dodecyltrimethylammonium bromide (DTAB), a cationic surfactant (See their structures in the Supplementary Information, SI, in Figure 1 SI). This system was here studied by a battery of techniques: surface tension, static (SLS) and dynamic light scattering (DLS), electrophoretic mobility, and bromide-ion – selective electrode. The mixed micelles' properties have been analyzed using two thermodynamic approaches: the Regular Solution Theory (RST, or Rubingh's approach) [16,17], and the Equation-Oriented Mixed Micelle Modellisation (EOMMM)[18]. These procedures give both the mixed micelle composition and thermodynamic properties such as the energy of intra-micellar interaction between components and their intra-micellar activity coefficients, which permit a better comprehension of what occurs inside the mixed system. On the other hand, we have experimentally obtained the composition of the mixed micelles at concentrations of about 10 times the CMC's. Although these values are not strictly comparable with those given by the thermodynamic approaches, that apply at the CMC, they allow us to evaluate the thermodynamic models and interpret the experimental findings.

## 2. MATERIALS AND METHODS

### 2.1 Materials

Dodecyltrimethylammonium bromide (DTAB) was from Sigma-Aldrich (>99%) and used as purchased. Polyethylene glycol p-(1,1,3,3-tetramethylbutyl)-phenyl ether (Triton X-100, TX100, Mw = 647gmol$^{-1}$) was obtained from Sigma-Aldrich and used as purchased. Stock solutions of DTAB and TX-100 were prepared using ultra-pure water (Milli-Q, Millipore system). Then, appropriate amounts of stock solutions were mixed and diluted to obtain the desired composition and concentration solutions. Mixtures having $\alpha_{DTAB}$ = 0 (pure TX100), 0.125, 0.25, 0.375, 0.5, 0.625, 0.75, 0.875, and 1 (Pure DTAB). Here $\alpha_{DTAB}$ is the mole fraction of DTAB in the whole surfactant mixture, without considering the solvent, i.e, $\alpha_{DTAB}$ + $\alpha_{TX100}$ = 1.

For dynamic light scattering experiments, the solutions were filtered three times through 220nm PDVF Millex filters from Millipore and let 24 hours to allow degasification.

## 2.2 Methods

All measurements were performed at 25.0 ± 0.1 º C.

Surface tension measurements were performed with a manual Krüss Tensiometer with a platinum duNoüy ring.

The refractive index increment at different concentrations was measured with a Phoenix Precision Instruments Co. differential refractometer with a controlled temperature cell jacket. The light source is a mercury lamp with filters to select the wavelength ($\lambda$ = 546 nm). The apparatus was calibrated with KCl solutions.

Dynamic light scattering (DLS) and electrophoretic mobility measurements were performed with a Malvern Zeta Ziser Nano ZSP with a He-Ne laser ($\lambda$ = 633 nm). Both DLS and electrophoretic mobility measurements were taken for total concentration of surfactant equivalent to ten times the CMC. DLS measurements relates the fluctuations in time of scattered light to the translational diffusion coefficient (*D*) [19], which may be related to the micelle hydrodynamic diameter, $d_H$. (see details in the SI, point 2.1). The temperature was controlled (± 0.1 °C) using the instrument own system.

The electrophoretic mobility (*u*) measurement is based in the laser Doppler velocimetry method with Phase Analysis Light Scattering (PALS) in order to obtain the electrophoretic velocity of the colloidal particles, *v*, and from it the mobility, *u=v/E*, E being the applied electric field. With *u* the zeta potential ($\zeta$) can be calculated by means of the Henry equation and Smoluchowsky approximation, $\zeta = \eta\,(u/\epsilon)$, where $\eta$ and $\epsilon$ are the solvent viscosity and permittivity respectively. Each mobility value was obtained as an average of several measurements, according to Malvern´s proprietary "Quality Factor" statistical criterion [20]. The total charge of the micelle is obtained with:

$$q = 6\pi\eta Ru \qquad (1)$$

R being the micelle radius.

The micellar mass was determined by static light scatteribng (SLS) with a Malvern Autosizer 4700 (laser OBIS Coherent 20 mW, λ= 514 nm) as a function of concentration and at a scattering angle of 90°, with a pinhole aperture of 300 μm. The temperature was controlled by the instrument system combined with a Lauda Alpha thermostatized circulating water bath. Static light scattering experiments in micelles can be interpreted from the Rayleigh's equation applied to particles smaller than light's wavelength [21] (see details in the SI, point 2.2).

Potentiometric measurements were performed with a Metrohm bromide ion-selective electrode, with a saturated calomel electrode as reference. Electric potential was read with a Titrino titrator, also from Metrohm.

## 3. THEORETICAL MODELS

Clint's model[22] relates the critical micelle concentration of a surfactant mixture, $CMC_M$, with the mole fraction in the mixture of components $i$, $\alpha_i$, and their pure critical micelle concentration, $CMC_i$:

$$CMC_M = \left[\frac{\alpha_1}{CMC_1} + \frac{\alpha_2}{CMC_2}\right]^{-1} \qquad (2)$$

Based on a simple phase separation model for micellization. Here $CMC_M$ is the value expected if the system behaves as an ideal mixture. The composition of the mixed micelle for component 1, $x_1$, is given by, $X_1 = \frac{\alpha_1 c - c_1^m}{c - c_1^m - c_2^m}$, where $c_i^m$ is the free monomer concentration of the i component. The mole fraction of component 1 in the micelle is defined as $X_1 = n_1/(n_1 + n_2)$, where $n_1$ and $n_2$ are the number of molecules of components 1 and 2 in a micelle. Even though Clint's model for the ideal mixed micelle solutions is appropriate only for very few systems, it has been used often as a way of analysing the deviation of a mixed system from the ideal behaviour[23].

The Regular Solution Theory (RST) or Rubingh's model[16] is the first model developed for non-ideal systems. It is based on a regular solution approach to the treatment of non-ideal mixing and due to its simplicity, it has been the mainly used approach, even after the development of more complex models. The non-ideality is introduced with the inclusion of the intra-micellar activity coefficients $\gamma_i$, into the equation relating the critical micelle concentration of the mixed system ($CMC_M$) and that of the i pure components ($CMC_i$):

$$CMC_M = \left[\frac{\alpha_i}{\gamma_i CMC_i} + \frac{\alpha_j}{\gamma_j CMC_j}\right]^{-1} \qquad (3)$$

In this model, for a binary solution:

$$\gamma_{1,M} = exp(\beta_M X_2^2) \; ; \quad \gamma_{2,M} = exp(\beta_M X_1^2) \qquad (4)$$

Where $X_i$ is the molar fraction of the surfactant i in the micelle (because the micelles composition could be different from bulk composition $\alpha_i \neq X_i$), and $\beta_M$ is an interaction parameter in $k_B T$ units, $k_B$ and T being the Boltzmann constant and the absolute temperature. The $\beta_M$ parameter can be interpreted in terms of the excess Gibbs free energy of mixing.

$$\beta_M = N_A(W_{11} + W_{22} - W_{12})/RT \qquad (5)$$

Here, $W_{12}$ is the energy of interaction between the surfactant molecules in the mixed micelles and $W_{11}$ and $W_{22}$ are the energies of interaction of surfactant molecules in an one single surfactant micelle, and R the gas constant. The parameter $\beta_M$ is determined from the following expressions, from experimental $CMC_1$, $CMC_2$ and $CMC_M$ values:

$$\beta_M = \frac{\ln(\alpha_1 CMC_M/X_1 CMC_1)}{X_2^2} = \frac{\ln(\alpha_2 CMC_M/X_2 CMC_2)}{X_1^2} \qquad (6)$$

Since there only the CMC's and the total system composition are known, the system formed by equations 3-6 is solved numerically for $\beta_M$ and $X_i$. for each experimental point (i.e., for each $\alpha_{DTAB}$) The $\beta_M$ quantitatively captures the extent of nonideality. The larger the negative values of $\beta_M$, the stronger the attractive interactions between the two different surfactants molecules. Repulsive interactions yields a positive $\beta_M$ value, whereas null $\beta_M$ indicates an ideal mixture. The first step in appling the model is to obtain the values of $X_1$ and $X_2$, by numerically solving equation 7, which relates the critical micelle concentration of the mixture and that of the pure surfactants to the $\alpha_1$. It is solved for each $\alpha_1$ value to obtain $X_1$ [17],

$$\frac{(X_1)^2 ln\left(\frac{\alpha_1 CMC_M}{X_1 CMC_1}\right)}{(1-X_1)^2 ln\left(\frac{(1-\alpha_1)CMC_M}{(1-X_1)CMC_2}\right)} = 1 \qquad (7)$$

The ideal composition of mixed micelles can be obtained with the Motomura and Aratono equation[24].

$$X_1^{id} = \frac{\alpha_1 CMC_2}{\alpha_1 CMC_2 + \alpha_2 CMC_1} \qquad (8)$$

The RST has received severe criticism[25,26]. In particular, its extension to multicomponent surfactant mixtures gives completely unrealistic results[27]. Since it is based in the regular solution theory, it is supposed that the energy if introducing a molecule of surfactant 1 in a micelle of pure surfactant 2, $W_{12}$, is equal to that of introducing a molecule of 2 in a micelle of pure 1, $W_{21}$, i.e., that the system is thermodynamically symmetric. This situation is very improbable. Moreover, in some cases it is not possible to resolve Equation (7) to obtain $X_i$.

The *Equation Oriented Mixed Micellization Model* (EOMMM) is a new approach based on Equation Oriented Optimization and Margules asymmetric formulations[28] contemplating both symmetric and asymmetric thermodynamic behaviors since the symmetric formulations are a particular case of the asymmetric ones, which is not restricted to the number of components and guarantees the applicability of the Gibbs-Duhem relation[18] . (For details see S.I. for *EOMMM*). The Equation Oriented Optimization simultaneously solves a system of equations in order to find the minimum/maximum of an objective function subject to a set of constraints. The EOMMM finds the Margules parameters (see S.I. Point 2.3) and the micelle compositions that globally minimize the total free energy of micellization. It has been recognized as main drawback that the original RST, and thus its multicomponent extension (MRST), deals with ionic surfactants as non-dissociated components. However, the EOMMM contemplates the dissociation of ionic surfactants through the *r* parameter and proper expressions for the activities of each component in the micelles. Thus, EOMMM can be employed for non-ionic or ionic surfactants, with or without the presence of supporting electrolyte. The EOMMM eliminates the assumption of interaction symmetry, i.e. $W_{12}$ and $W_{21}$ are no restricted to be identical. At difference of the original RST, the solution is obtained taking into account all the data simultaneously. The model computes the Gibbs free energy of mixing and that in excess ($\Delta G_{mix}$ and $\Delta G_{mix}^{exc}$) of mixing, the the intra-micellar activity coefficients, the values of $W_{12}$ and $W_{21}$, and micelle composition. Moreover, the procedure may be extended to multicomponent surfactant mixtures. The procedure is explained in detail in the SI, point 2.3.

For the EOMMM analysis the pure DTAB micelle ionization degree was taken from literature, and an average value of 0.260 ± 0.004 was used[29]

## 4. RESULTS

In order to experimentally determine the composition of the micelles for each mixture, the micellar mass of the mixed micelles was determined by SLS (see SI). Now, since Triton X-100 is a non-ionic surfactant, the charge of mixed micelles is due to the DTA$^+$ and Br$^-$ ions content. To determine if bromine counterions are condensed on the mixed micelles, a Br$^-$ - ion selective electrode was used. As it can be seen in Figures 2 SI and 3 SI, for $\alpha_{DTAB} < 0.75$, there is no capture of couterions and then the charge of the micelle is equal to their DTA$^+$ -ions content.

With the micelle hydrodynamic radius, the micellar mass and the micelle electrophoretic mobility for each mixture composition, an estimation of the micelle composition may be performed. The micelle charge may be obtained from the electrophoretic mobility of micelles. Since at most for $\alpha_{DTAB} \leq 0.75$, micelles are fully ionized, the charge of micelles (in terms of the elementary charge e) is equal to the number of DTA$^+$ ions included in the micelle ($n_{DTA+} \approx q$). Then, using the micellar mass, the composition of mixed micelles was computed as:

$$M_{micelle} = n_{DTA^+}M_{DTA^+} + n_{TX-100}M_{TX-100} \quad (9)$$

Where $M_i$ is the molar mass of component i in the mixed micelle. The composition of the mixed micelle was then computed as $X_{DTA+} = n_{DTA+}/(n_{DTA+} + n_{TX-100})$, and its aggregation number $n = n_{DTA+} + n_{TX100}$.

For $\alpha_{DTAB} > 0.75$, $n_{DTA+}$ was estimated using the surface areas and partial molar volumes of the components and micelles:

$$V_M = 4\pi(d_h/2)^3/3 = n_{DTA}V_{DTA} + n_{TX100}V_{TX100} \quad (10)$$

$$A_M = 4\pi(d_h/2)^2 = n_{DTA}A_{DTA} + n_{TX100}A_{TX100} \quad (11)$$

Where $d_h$ is the hydrodynamic diameter of micelles, $A_M$ and $V_M$ are the area and volume of a mixed micelle, and $A_i$ and $V_i$ the area per polar head group and molecular volume of component i. The values here used are $A_{DTA} = 0.3752$ nm$^2$, $A_{TX100} = 2.668$ nm$^2$, $V_{DTAB} = 0.1617$ nm$^3$ and $V_{TX100} = 4.157$ nm$^3$. All values were computed using the volume and area of the pure surfactant micelles divided by their aggregation numbers. Assumptions subjacent of the computations are that micelles are spherical and that the molecular volumes and areas do not change when passing from pure surfactant to mixed surfactant micelles. In particular, TX100 include a large amount of water molecules, at 25 ºC it amounts $\delta$ = 0.3697 $g_{water}/g_{TX100}$.[30], i.e., about 14 water molecules per micelle. The method used here incudes the hydration water in the computed $V_{TX100}$ value. The TX100 micellar partial molar volume is $PMV_{TX100}$ = 587.06 cm$^3$/mol[30] which gives a molecular volume of micellised TX100 of 0.8864 nm$^3$. Although the PMV is generally taken as the volume of the molecule, this is not its correct interpretation. The $PMV_{TX100}$ indicates how the inclusion of TX100 molecules affect the total volume of the solution, including structure making and structure breaking effects and electrostriction. For instance, some ions have negative PMV in water. Then, the comparison between the $PMV_{TX100}$ and the micellised molecular volume computed from the micellar kinetics entity must be taken with caution.

The CMC for each mixture composition was determined by two different methods: surface tension and static light scattering (see examples of the experimental results in SI Figures 4 SI and 5 SI). The experimental values are shown in SI, Table SI-I. The average value for the TX100 CMC was (2.07 ± 0.23)x10$^{-4}$ mol.dm$^{-3}$, that of literature is 3.1x10$^{-4}$ mol.dm$^{-3}$ [31], 3.31x10$^{-4}$ mol.dm$^{-3}$ [15] or (2.550 ± 0.015)x10$^{-4}$ mol.dm$^{-3}$ as an average of several literature values [29]. That of DTAB CMC is 0.0144 mol.dm$^{-3}$, that of literature is 0.015 mol.dm$^{-3}$ [15,32], depending on the experimental method used the cmc falls between 0.014 and 0.016 mol.dm$^{-3}$ [33].

Results are shown in Figure 1. The Clint ideal CMC (Equation 2) is also represented as a continuous line. It can be seen that the experimental results are close to the ideal prediction.

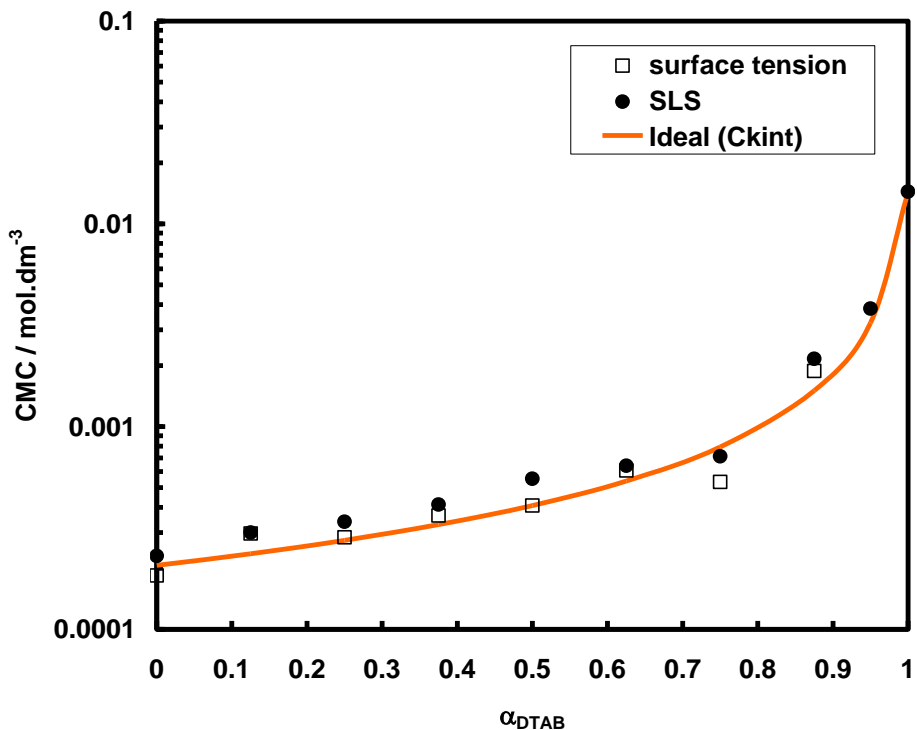

Figure 1. Critical micelle concentration dependence on the mole fraction of DTAB in the whole mixed system obtained from surface tension (□) and static light scattering (•) measurements. The continuous red line corresponds to the ideal behaviour predicted by the Clint model.

In a literature work[15] the CMC of DTAB-TX100 mixed micelles dependence on $\alpha_{DTAB}$ was also almost ideal, while that of TTAB-TX100 and CTAB-TX100 systems were not, showing experimental values below those predicted by Clint equation

The micelle mass (M) determined by SLS is plotted in Figure 2 as a function of $\alpha_{DTAB}$. An example of the Debye plots can be seen in Figure 6 SI. The values may be seen in Table SI-II. From literature, M = 66700 Da for pure TX100 micelles in water at 25 ºC[34]. Other literature values are M = 58000 Da[35], 87930 ± 740 Da as an average of seven values summarized in Robson and Dennis work[36]. For DTAB M = 20900 Da[29] or 15500 Da[37].

In figure 2 we observe that M decreases as $\alpha_{DTAB}$ increases. This may be due to an increased repulsion among the micellised molecules when the DTAB content increases. There is a change in the M dependence on the mixture composition at $\alpha_{DTAB}$ = 0.5.

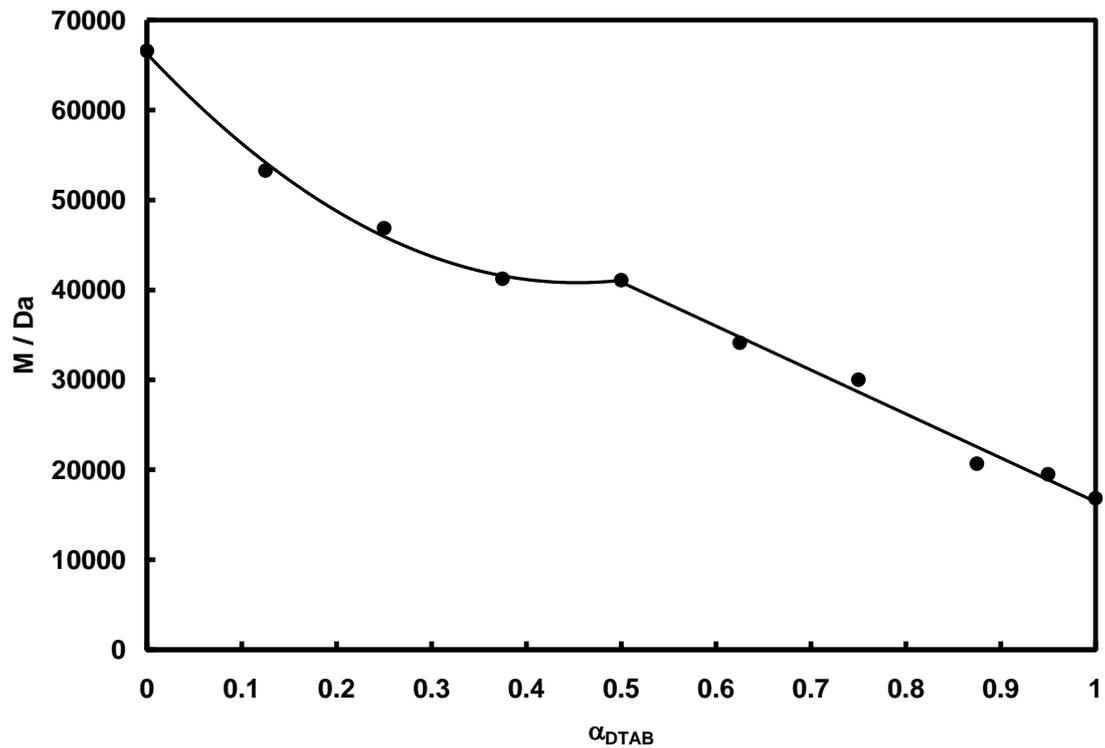

Figure 2: Micelle masses obtained from the Debye's equation versus the mixture composition. Lines are eye guides.

The static light scattering experiments also gives the second virial coefficients ($A_2$) measured for each mixture, which are shown in Figure 3.

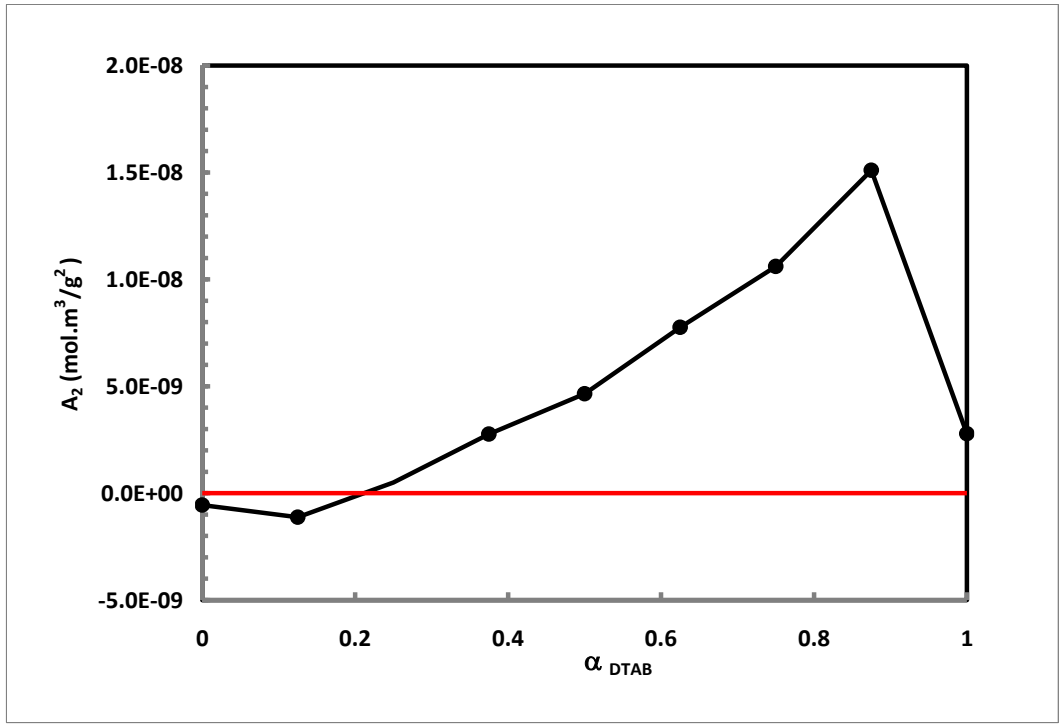

Figure 3: The second virial coefficient from the Debye static light scattering plot, as a function of the overall mixture composition. The red straight line indicates zero.

Figure 4 shows the hydrodynamic diameter of micelles for different mixtures, measured by dynamic light scattering (DLS). The size of micelles decreases with the increase of $\alpha_{DTAB}$. The slope of this decrease changes at $\alpha_{DTAB}$ = 0.25 and 0.85.

The values for $\alpha_{DTAB}$ = 0 and $\alpha_{DTAB}$ = 1 are in accordance with literature[38][39]. For $\alpha_{DTAB}$ = 0 (pure TX100) Bulavin *et al.*[40] measured by small-angle neutron scattering a constant characteristic diameter of 7.4 nm below 0.0096 mol.dm$^{-3}$. Mandal *et al.* [30] proposed an oblate ellipsoidal micelle for TX100, with an hydrodynamic radius of 3.962 nm ($d_h$ = 7.924 nm), a gyration radius $R_G$ = 3.343 nm and an equivalent sphere radius $R_o$ = 3.610 nm. The oblate ellipsoid semiaxes are *a* = 5.131 nm and *b* = 1.796 nm.

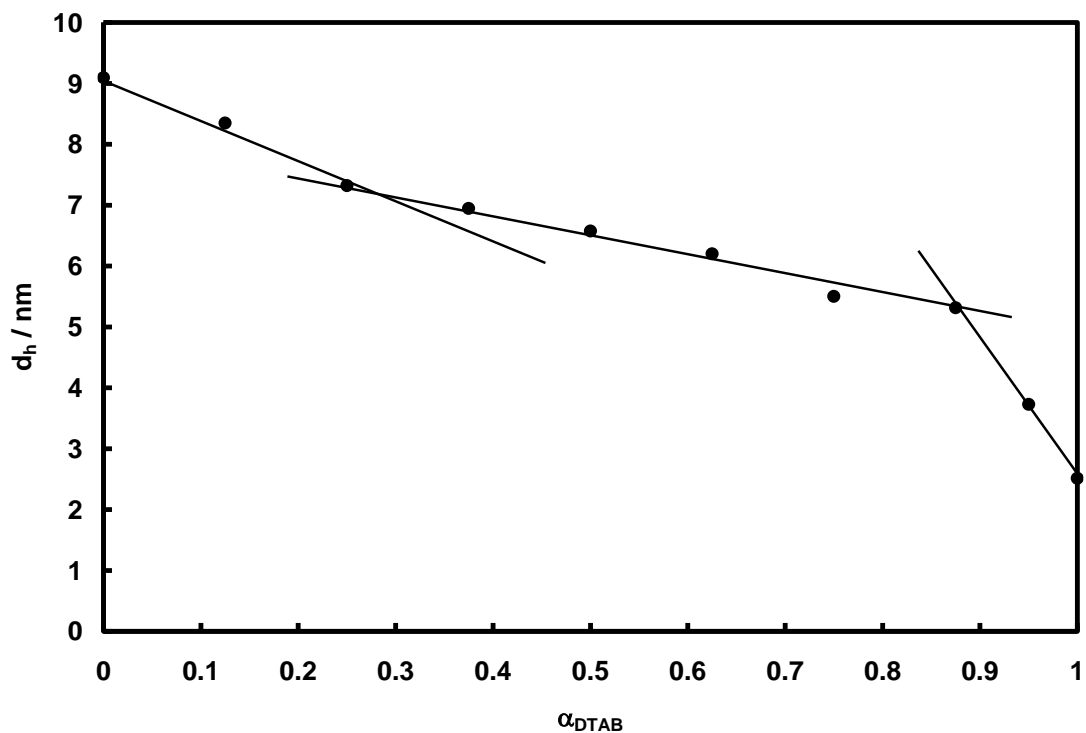

Figure 4 Variation of the hydrodynamic diameter for micelles for different total molar fraction of DTAB from DLS

Figure 2 SI shows the dependence on the total concentration C of the potential (E) of the Br$^-$-ion-selective electrode against the saturated calomel electrode (SCE) for the mixtures having $\alpha_{DTAB}$ = 0.25, 0.50, and 0.75. Since there is not a break at the CMC, it may be concluded that these mixed micelles do not capture counterions at their surface. This behaviour was observed for all systems with $\alpha_{DTAB} \leq 0.75$, while for higher DTAB proportions micelles capture some counterions, as it can be seen in Figure 3 SI, although this capture is not so high as in pure DTAB micelles[41].

Then, at least for $\alpha_{DTAB} \leq 0.75$, the micelles are completely ionized, i.e., they do not have bromide ions included in the kinetic unit, due to their very low surface potential.

Fluorescence anisotropy studies on the aklyltrimethylammonium bromides-TX100 gave information about microviscosity in the aggregates interior. Literature measurements indicated that the structure of micelles is less tightly packed in mixed aggregates than in pure TX100 ones[15]. This fact allows the positively charged heads of the DTAB to be far apart, causing a low surface charge density. Thus the counterions should not be electrostatically attached to the micelles. However, in another mixed system, DTAB-sodium undecenoate (SUD) it was found, that even micelles with compositions that had a SUD content enough to make them negatively charged, there was aggregation of bromide ions to the micelle surface[42], and the same was found in a computer simulation of the same system[43]. This adsorption was attributed to van der Waals adsorption of Br$^-$ ions to the micelle-solution interface, due to the high polarizability of this ion. This situation may be hindered by the broad, strongly hydrated polyoxyethylenic shell of micelles predominantly formed by TX100.

In figure 5 the measured micellar electrophoretic mobilities ($u$) is plotted as a function of $\alpha_{DTAB}$. Both electrophoretic mobility and zeta potential measurements are summarized in Table SI-III and Figure 7 SI. The electrophoretic mobility increases linearly until $\alpha_{DTAB}$ = 0.75, where the slope of the data changes abruptly. This behaviour is similar to that observed in the micelle size vs. $\alpha_{DTAB}$ plot (figure 4).

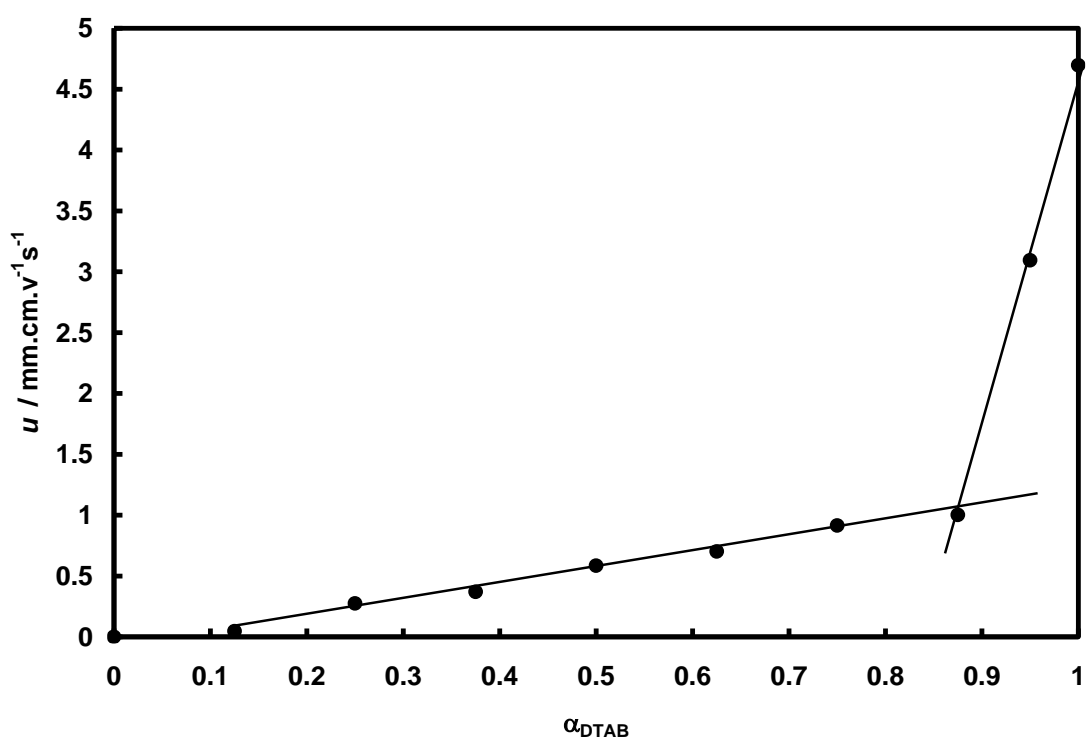

Figure 5 Electrophoretic mobility of mixed micelles vs $\alpha_{DTAB}$. Lines are eye guides.

With the above data, using Equation (9), the composition of mixed micelles was computed and results are presented in Table I It must be taken into account that the data employed in this computation are obtained in different conditions. The micellar mass is obtained at the CMC, while the hydrodynamic radius and the zeta potential were obtained at 10CMC. Taken into account the observations also given in the theory section about the procedure *ut supra*, results must be seen with caution.

Table I.

Experimental hydrodynamic diameter ($d_h$) and computed micelle composition and aggregation number of micelles at different mixture overall composition.

| $\alpha_{DTAB}$ | $X_{DTAB}$ | $N_{agg}$ | $d_H$ / nm |
|---|---|---|---|
| 0 | 0 | 103 | 9,3 |
| 0,125 | 0.003 | 82 | 8,3 |
| 0,25 | 0.012 | 73 | 5,2 |
| 0,50 | 0.040 | 65 | 6,6 |
| 0,75 | 0.072 | 48 | 5,5 |
| 0,95 | 0.63 | 45 | 3,8 |

Some literature values of $N_{agg}$ for TX100 micelles in water are 140 [44], 111 [34] and 135 [45]. For DTAB micelles, the measured aggregation number found in literature are in the range of $N_{agg}$ = 40 – 73 [33][46][47]

The application of the RST to the CMC data was quite unsuccessful, because Equation (7) only gave results for two points, that are presented in Table II. For the other points, equation (7) cannot be solved numerically (it didn't converge), and then $X_{DTA+}$ could not be obtained.

Table II

Results of the Rubing's method in the system. $\Delta G_{mix}^{exc}$ is the excess Gibbs free energy of micellization.

| $\alpha_{DTAB}$ | $X_{DTAB}$ | $\gamma_{DTAB}$ | $\gamma_{TX100}$ | $\beta_M$ $k_BT$ | $\Delta G_{mix}^{exc}$ RT |
|---|---|---|---|---|---|
| 0,75 | 0,173 | 1,07 | 0,21 | -2,25 | -0,322 |
| 0,95 | 0,187 | 0,98 | 1,36 | 0,46 | 0,07 |

It must be realized that this model gives the properties of the mixed micelles at the CMC, and then they are not strictly comparable with the compositions found with the experimental procedure.

In a literature study [15] the Rubingh's intramicellar inrteraction parameter $\beta_M$ for the DTAB-TX100 system varied from one point to other, whilst for the TTAB-TX100 and CTAB-TX100 systems it remained rather constant. These authors concluded that the RST cannot be applied to the DTAB-TX100 mixtures. In all these systems the $\beta_M$ values became more negative as $\alpha_{TX100}$ decreased, and this trend was more sharp in the DTAB-TX100 system than in the other cationic homologues. This is an indication of the thermodynamic asymmetry of the system, i.e., that $W_{12} \neq W_{21}$. Moreover, when applying the Maeda formulation[48] to determine the excess free energy of mixing, Carnero Ruiz and Aguiar[15] found that the parameter $B_2$ = +1.23. Its interpretation is the same as $W_{12}$, i.e, it is related to the standard free energy upon the replacement of a nonionic monomer with an ionic one (see below the EOMMM results).

Figure 6 shows the EOMMM results. As it can be seen from the $\Delta G_{mix}^{exc}/RTX_{TX100}X_{DTA}$ line, the system is extremely asymmetric, which may be the cause of the failure in the application of the RST, which assumes symmetry. The values of $W_{12}$ and $W_{21}$ are very different. $W_{12}$ represents the energy to introducing a DTAB molecule in a pure Triton X-100 micelle, and amounts +4.04$k_B$T, i.e., it indicates a repulsive interaction. Conversely, $W_{21}$, the energy of introducing a Triton-X100 molecule in a pure DTAB micelle, is –14.02$k_B$T, i.e, indicating a strong attractive interaction.

As it can be seen, the excess free energy of mixed micellization is relatively low, as expected from the almost ideal dependence of the CMC on $\alpha_{DTAB}$.

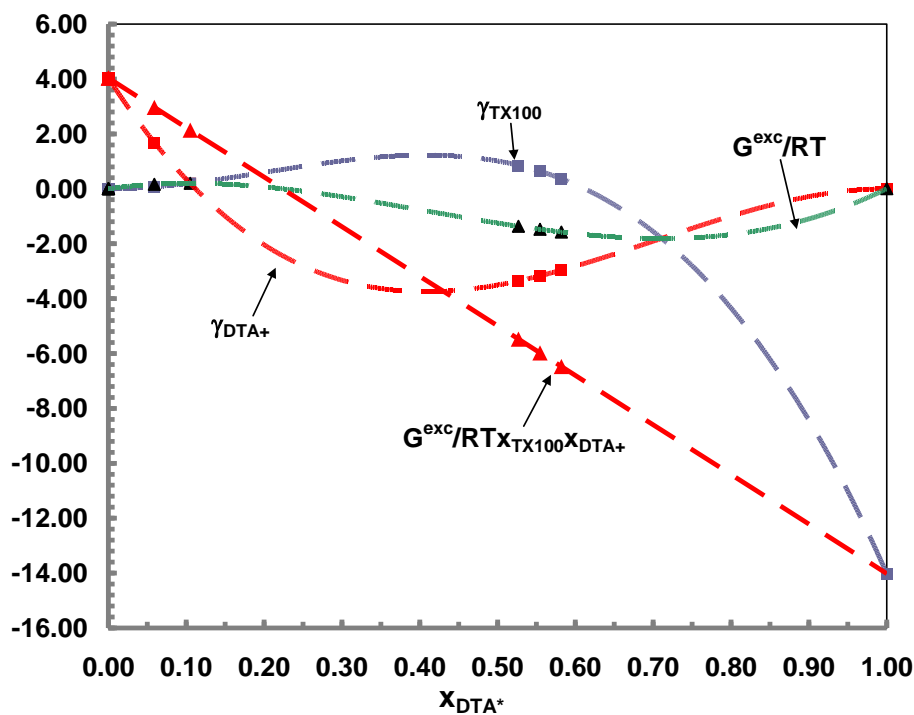

Figure 6: The results of the application of the EOMMM to the DTAB-Triton X-100 mixed system.

Figure 8 SI shows the CMC values fitted by the EOMMM for different $\alpha_{DTAB}$, in comparison with the experimental and ideal (Clint equation) ones. The fitting is very good.

Figure 7 shows the activity coefficients of the components in the mixed micelles, and Figure 8 the values of $X_{DTA+}$ as a function of $\alpha_{DTAB}$.

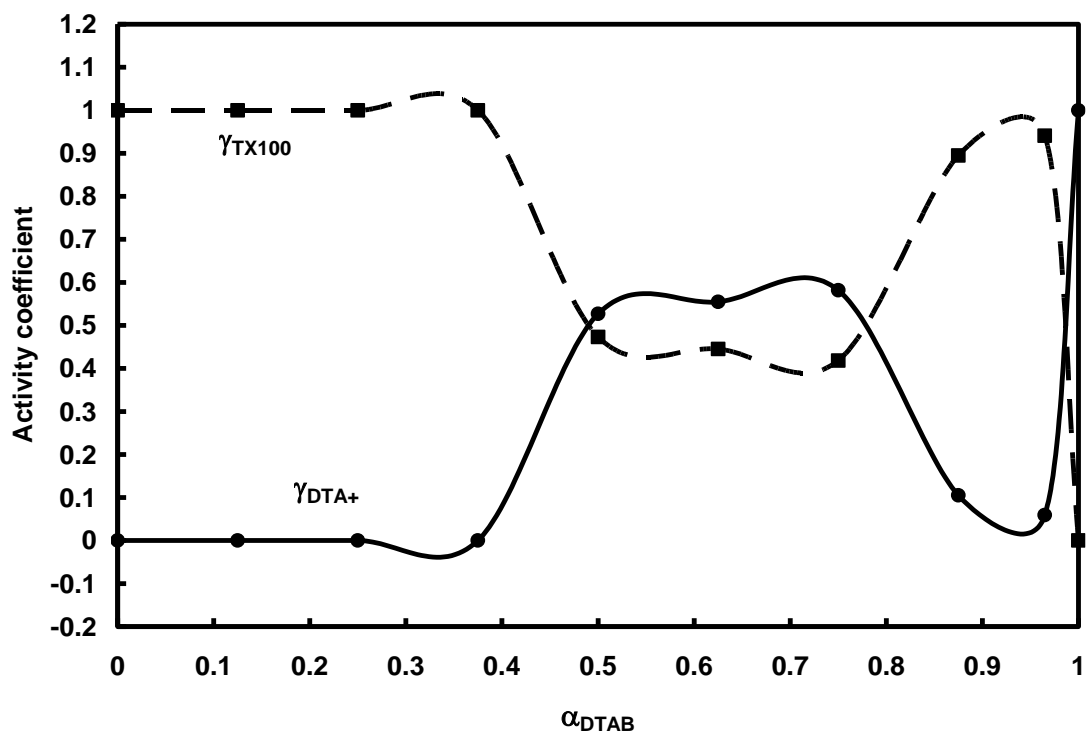

Figure 7: the intramicellar activity coefficients of DTA[+] and Triton X-100 as a function of the mixture composition.

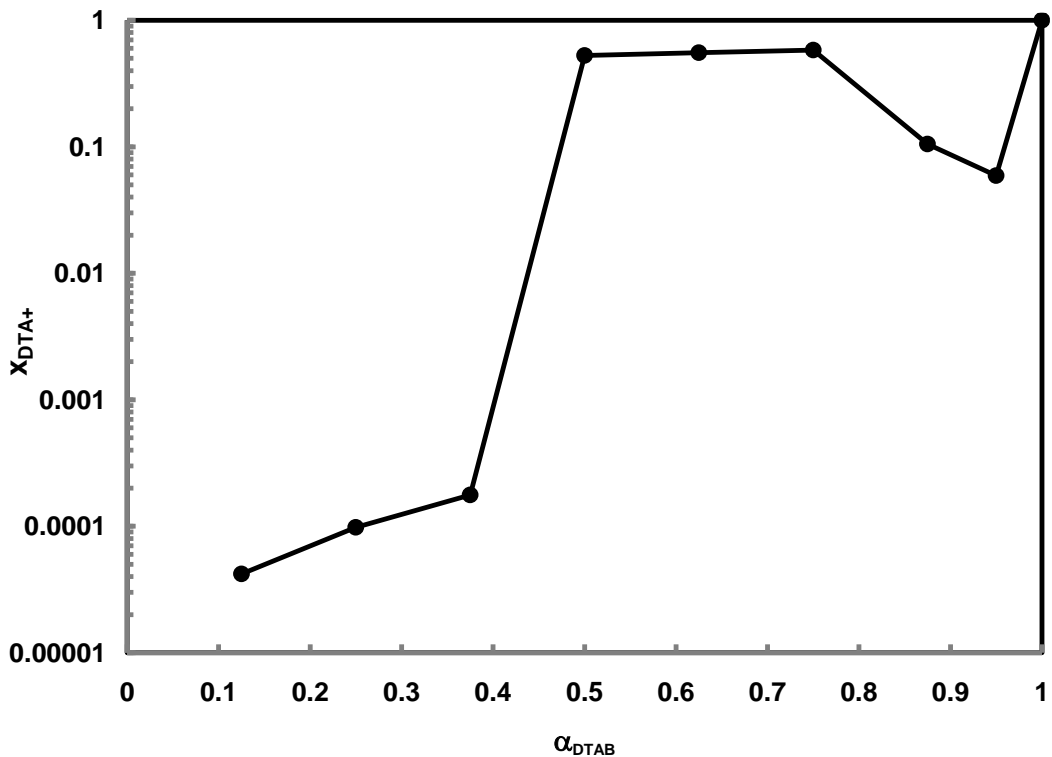

Figure 8: the composition of mixed micelles at the CMC vs. the mole fraction of DTAB in the mixture, accordingly the EOMMM.

Carnero Ruiz and Aguiar[15]. also found that in the system DTAB-TX100 system, when $\alpha_{DTAB}$ is low, the content in cationic surfactant in mixed micelles is very low also, and that the DTAB content in the mixture increases its inclusion in the aggregates becomes significant.

Figure 9 shows the concentration of Triton X-100 at the mixture CMC, computed as [Triton X-100] = $\alpha_{TX100}$CMC.

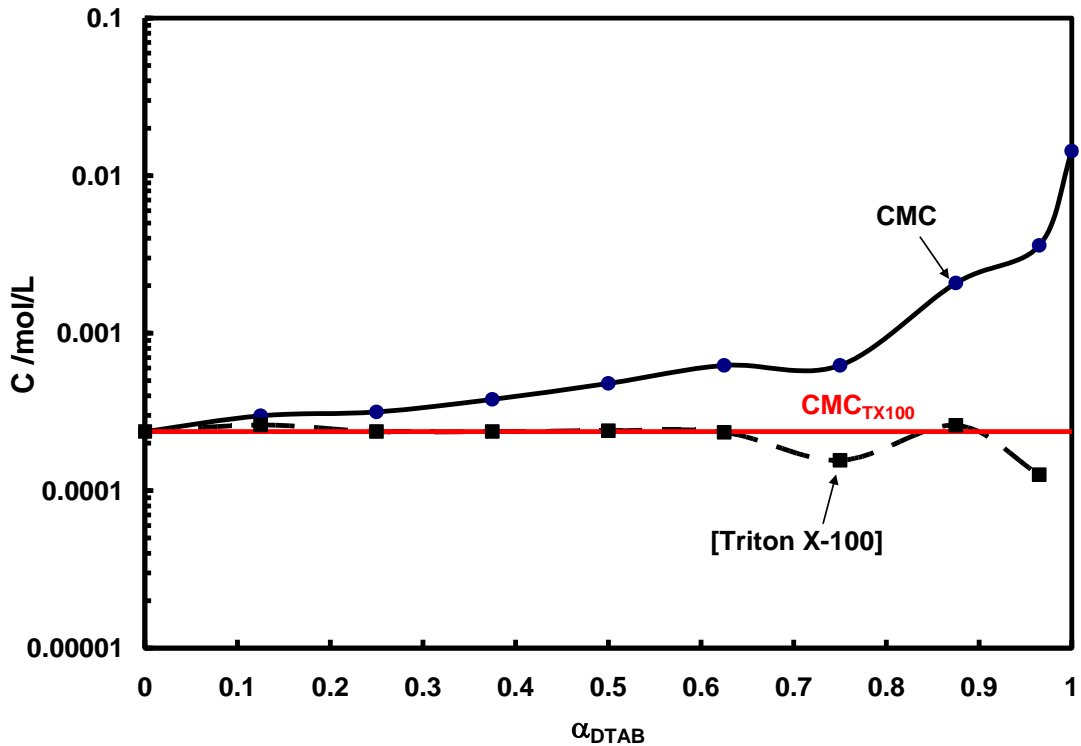

Figure 9: The concentration of Triton X-100 (■) at the mixture CMC (●), and the pure Triton X-100 CMC (----), as a function of the surfactant mixture composition.

## 5. DISCUSSION

The CMC values determined from surface tension and SLS are in agreement (Figure 1), and their dependence on $\alpha_{DTAB}$ is almost ideal, which is astonishing in view of the very different nature of both surfactants. Triton X-100 has a branched hydrophobic chain, a bulky aromatic ring and a highly hydrated polyoxyethylene chain, while DTAB has a straight hydrocarbon chain and a relatively small ionic head group. In Figure 6 it can be seen that the excess free energy of mixed micellization ($G^{exc}/RT$) is nearly zero, indicating that the interaction is almost ideal, too.

From Figure 6 it is also evident that, in spite of this almost ideal behavior, the system is very asymmetric. This may be the cause of the failure in the application of the Rubing's procedure, which assumes symmetry. Moreover, the very strong difference between $W_{12}$ and $W_{21}$ (+4.04$k_B$T and –14.02$k_B$T, respectively) makes very difficult to obtain an unique value of $\beta_M$. In fact, the only two values obtained for the intramicellar interaction parameter (+0.46$k_B$T and -2.25$k_B$T) are too different to make an average with any significance. This suggests that the values of $X_{DTAB}$ obtained from the RST are scarcely reliable.

The values of $W_{12}$ and $W_{21}$ (+4.04$k_B$T and –14.02$k_B$T, respectively) may be interpreted as follows: The inclusion of a DTA$^+$ ion in a pure TX100 micelle is energetically unfavorable, because it introduces a charge in a nonionic micelle. The inclusion of other ionic surfactant units will increase the inter-head group repulsion, which is in conflict with micellization. Then, there is some repulsion between the two components. This also may be the cause of the low $X_{DTA+}$ values for low $\alpha_{DTAB}$ mixtures (Figure 8).

TX100 minimal micelles are adequately represented as hard spheres[49]. It was proposed[36][50][51] that there is not a sharp boundary between the hydrophobic interior and the polyoxyethylene chain shell of TX100 spherical micelles. In those articles, authors have also suggested that the first oxyethylene groups of the alkylphenol and some TX100 molecules are contained in the hydrophobic core.

Pirene fluorescence has been used in alkytrimethylammonium bromide-TX100 mixed micelles to study the micropolarity of aggregates[15]. Pyrene locates near the surface of the hydrocarbon core of micelles, and the determinations were made well above the CMC. The micropolarity of micelles decreases from pure TX100 micelles when the DTA$^+$ content in aggregates increases. This was attributed to an increase in ion-dipole interactions between trimethylammonium and the oxyethylene groups. This in turn causes a partial dehydration of the polyoxyethylenic chains and a reduction in the micelle volume with increasing $\alpha_{DTAB}$, as observed in this work (Figure 4). The inclusion of the cationic surfactant into mixed micelles produced more crowded aggregates with a more dehydrated structure. Mixed micelles had a less ordered structure than that of pure TX100 ones[15]. Dehydration is an energy consumer process which may contribute to the TX100 micelles reluctance to include DTA$^+$ ions.

Accordingly Yuan *et al.*[52], the $\alpha$-methylene group of CTAB is in the near vicinity of the phenoxy ring of TX100. The trimethylammonium group of CTAB locates between the first oxyethylene group next to the phenoxy ring of TX100 and the methyl terminal group of the hydrophobic chain of CTAB is close to those of the nonionic surfactant. CTAB and TX100 are uniformly distributed in the mixed micelles. This latter conclusion indicates that the excess entropy of mixed micellization will be near zero. This is one of the assumptions of the RST procedure, which may be an explanation of the nearly ideal mixture CMC behavior of the system.

Thus, the hydrophobic core of pure TX100 micelles is not completely apolar, and this does not favor the inclusion of the hydrocarbon chain of DTA$^+$ ions. This may be the cause of the initial repulsion and the positive value of $W_{12}$.

On the other hand, the inclusion of a TX100 molecule in pure DTAB micelles ($W_{21}$) is very favorable because it introduces a bulky, uncharged headgroup between charged groups, thus reducing their mutual electrostatic repulsion energy. In TX100-hexadecyltrimethylammonium bromide (CTAB) micelles, the trimethylammonium groups are situated facing the aryl groups of TX100, probably interacting with their π-electrons [4]. Similar conclusions were formulated in TX100-sodium dodecylsulfate (SDS) mixed micelles[53].

The $X_{DTAB}$ values experimentally determined were computed with data of diverse origin and at different surfactant concentrations. In particular, DLS and electrophoretic mobility were measured at 10CMC, then the results cannot strictly be compared with those of both, the RST and the EOMMM procedures, which are computed at the CMC. Previous results in another system showed that the composition of the micelles changes strongly when increasing the total concentration of the surfactant. These conclusions were obtained from experiments without using any model[4][54]. As a consequence, we will use just the EOMMM results in the following discussion.

As it can be seen in Figure 4, there is a change in the dependence of the hydrodynamic diameter on the mixture composition at $\alpha_{DTAB} \approx 0.75$. There is an abrupt decrease in $d_h$ above this value. The electrophoretic mobility increases abruptly in the same region (Figure 5).

Although Fang et al.[4] did not explored all the composition range in TX100-CTAB mixtures, the behavior of the mixed micelles' diffusion coefficient showed the same tendency as our results, with a monotonically decrease indicating a reduction in size when the proportion of the ionic surfactant in the overall mixture increases.

The evolution of the micelle mass with $\alpha_{DTAB}$ is monotonically decreasing but there is a slight change above $\alpha_{DTAB} \approx 0.5$ (Figure 2).

Comparison with the evolution of $X_{DTAB}$ with $\alpha_{DTAB}$ from the EOMMM (Figure 8) indicates a sudden change in the composition of mixed micelles at $\alpha_{DTAB} \approx 0.5$.

Figure 7 shows that below $\alpha_{DTAB} \approx 0.5$ Triton X-100 acts as a solvent ($\gamma_{TX100} \approx 1$) whilst DTA$^+$ activity coefficient is near zero. Between $\alpha_{DTAB} \approx 0.5$ and $\alpha_{DTAB} \approx 0.75$ it seems that both surfactants form a mixture in which none of them acts as a solvent and the other as a solute, and above this latter composition, there is a new change in the interaction. These changes are reflected in the diverse properties here studied, as discussed *ut supra*.

As it was said above, Robson and Dennis[36] suggested spherical TX100 micelles, having some polyoxyethylene chains immersed into the hydrophobic micelle core. The inclusion of DTA$^+$ hydrocarbon tails may change this composition, and eventually the mixed micelle core becomes fully hydrophobic. This may occur at $\alpha_{DTAB} \approx 0.4$, causing the changes detected in Figure 7, with mutual solubility indicated by the almost equals activity coefficients.

Figure 9 shows that the mixture CMC remains close to that of the pure TX-100 along the composition range. This is a behaviour also found in TX100-CTAB mixtures, in which the presence of the cationic surfactant caused small perturbation to the micellization behaviour of TX100 [4]. Then, it is evident that TX100 micelles act as a rather ideal solvent for DTA$^+$ ions, at most up to $\alpha_{DTAB} \approx 0.40$.

From Figure 9 it also may be seen, mixed micelles form when the nonionic surfactant reaches its pure CMC, i.e., it may be interpreted that first micelles of Triton X-100 are formed and then they capture some DTA$^+$ ions. Apart of the energetically disfavored inclusion of the ionic surfactant, this may be also due in part to the very different CMC values of the mixture components.

The above results may be interpreted as follows: Triton X-100 incorporate DTA$^+$ ions, but the inclusion is initially difficult, showing some repulsion, as it can be appreciated with the $W_{12}$ value. This causes mixed micelles formation with small DTA$^+$ content, as it can be seen in Figure 8. The incorporation of DTA$^+$ molecules will increase the charge of micelles and reduce its diameter. This also increases the repulsion among micelles. Figure 3 shows the second virial coefficient from the Debye plot in SLS. Micelles very rich in Triton X-100 show small negative $A_2$ values, indicating an attractive interaction due to van der Waals interactions. With increasing DTA$^+$ content the electrostatic repulsion increases and then also the positive $A_2$ values do. When $\alpha_{DTAB} = 1$ there is a reduction of $A_2$ caused by the inclusion of counterions in the micelle Stern layer and the reduction of the Debye length caused by the high ionic concentration, because of the higher CMC of DTAB.

In micelles for $\alpha_{DTAB} \geq 0.75$ the DTA$^+$ content is high enough to capture some counterions, and the Triton X-100 content small enough to permit a reduction of the effective hydrodynamic diameter. This is because if the nonionic surfactant is not tightly crowded in the micelle, the polyoxyethylene chains may be folded instead of extended to the intermicellar solution. The partial dehydration of the polyoxyethylene chains may contribute to this. Similar conclusions were obtained in the TX100-CTAB mixed system[4]. Both effects increase the electrophoretic mobility and reduce the micelle mass.

Then, there is a change in the nature of micelles, from predominantly Triton X-100 micelles having solubilized some DTA$^+$ ions at $\alpha_{DTAB}$ below about 0.75 to predominantly DTAB micelles having solubilized Triton X-100 above that mixture composition, passing through micelles having apolar hydrocarbon core between $\alpha_{DTAB} \approx 0.40$ and 0.75.

Although discussed in other terms, Fang *et al.* [4] found some changes in the TX100-CTAB mixed micelles behaviour above and below about $\alpha_{CTAB} \approx 0.5$, related to changes in the interaction between the components of the system. As an example, the micelle composition behaviour when the total concentration is increased is the opposite for higher $\alpha_{CTAB}$ values (increasing with increasing concentration C) than for lower ones (decreasing with increasing C). Unfortunately, the region with $\alpha_{CTAB} > 0.75$ was not studied by these authors.

In mixtures of TX100-SDS, Zhang and Dubin[53] found evidence of a coexistence of two different mixed micelles: TX100 rich micelles with some solubilized DS$^-$ ions, and SDS-rich micelles with some solubilized TX100 molecules. This was attributed to the possibility of energetically equivalent micelles of different composition, because of the different form in which the inclusion of one component molecule or ion in the other component micelle affects $\Delta G_{mic}$.

Although based on a symmetric intramicellar interaction energy in mixed micelles, Barzikin and Almgren [55] have theoretically demonstrated that if this interaction is positive (as it occurs in mixtures of hydrocarbon-based with fluorocarbon – based surfactants) it is possible the formation of a two-phase micellar system, i.e., a mixture of coexisting mixed micelles having different composition, both having the same free-energy of mixing. In view of the values of $W_{12}$ and $W_{21}$ in the TX100-DTAB system, this may be the explanation of the phenomena here found. Although this is a speculation, it is possible that the formation of the coexisting micelles occur in the range $0.5 \leq \alpha_{DTAB} \leq 0.75$; and that for $\alpha_{DTAB} > 0.75$ one of the two kinds of micelles predominates and the other trends to disappear. Since the techniques here used (including the RST and EOMMM application) give average values for the different properties, there is necessary to use another procedure that permit clarify this point.

## 6. CONCLUSIONS

In this article, we presented a systematic study of the micelles formed by a mixture of a non-ionic (TX-100) and a cationic (DTAB) surfactant. Both from theoretical and experimental findings we arrive to the following concluding remarks,

- In spite of the very different molecular structure of the components and the very asymmetric thermodynamic interactions, the CMC of the mixtures have a nearly ideal behavior.
- The system is very asymmetric with the energy of introducing a $DTA^+$ ion into a TX100 micelle positive, meaning certain repulsion, while the introduction of a TX100 molecule in a pure DTAB micelles has a strong negative value, indicating attraction. This asymmetry is caused by the strong difference in surfactants structure.
- This asymmetry can be the cause of the failure in the application of the RST procedure to this system
- Both hydrodynamic diameters and electrophoretic mobilities change abruptly their dependence on $\alpha_{DTAB}$ at $\alpha_{DTAB} \approx 0.75$.
- The hydrodynamic diameter and the micelle mass decrease, and the electrophoretic mobility increases with increasing $\alpha_{DTAB}$, indicating a decrease in micelle size and an increase in micelle charge with the addition of DTAB.
- Below $\alpha_{DTAB} \approx 0.5$ micelles are mainly of TX100 acting as a solvent for some $DTA^+$ ions. Between this value and $\alpha_{DTAB} \approx 0.75$ there seems not to exist any difference in which is the solvent in the mixed micelle, and above this value there seems that an abrupt change in the system aggregates' structure occurs. These changes are reflected by different experimental results.
- On the basis of other TX100 mixed systems, there seems possible that two different micelles can coexist, at most above $\alpha_{DTAB} \approx 0.5$, the different results being an average of the properties of both types of micelles.

In a near future we will continue studying these surfactant mixtures focusing on structural changes as a function of micelle composition. In doing so we will introduce a very sensitive technique, electric birefringence [6], which could be capable of discerning the presence of different micelles (with different surface charges and membrane viscoelasticities), in order to explore the validity of the final statement of this conclusion about the coexistence of two types of mixed micelles.

**Acknowledgements**

This work was partially supported by grants PGI-UNS 24/F067 and 24/Q073 of Universidad Nacional del Sur, PICT-2013(D)-2070 and PICT-2016-0787 of Agencia

**References.**

[1]  M.J. Rosen, Surfactants and Interfacial Phenomena, 3rd ed., Wiley-Interscience, New Jersey, 2004.

[2]  D. Myers, Surfactant science and technology, J. Wiley, 2006.

[3]  M. Abe, J.F. Scamehorn, eds., Mixed Surfactants Systems., 2nd ed., Marcel Dekker, New York, 2005.

[4]  X.W. Fang, S. Zhao, S.Z. Mao, J.Y. Yu, Y.R. Du, Mixed micelles of cationic-nonionic surfactants: NMR self-diffusion studies of Triton X-100 and cetyltrimethylammonium bromide in aqueous solution, Colloid Polym. Sci. 281 (2003) 455–460. doi:10.1007/s00396-002-0797-6.

[5]  E. Guzmán, S. Llamas, A. Maestro, L. Fernández-Peña, A. Akanno, R. Miller, F. Ortega, R.G. Rubio, Polymer–surfactant systems in bulk and at fluid interfaces, Adv. Colloid Interface Sci. 233 (2016) 38–64. doi:10.1016/j.cis.2015.11.001.

[6]  H.A. Ritacco, Electro-optic Kerr effect in the study of mixtures of oppositely charged colloids. The case of polymer-surfactant mixtures in aqueous solutions, Adv. Colloid Interface Sci. 247 (2017) 234–257. doi:10.1016/j.cis.2017.05.015.

[7]  M.M.S. Lencina, E. Fernández Miconi, M.D. Fernández Leyes, C. Domínguez, E. Cuenca, H.A. Ritacco, Effect of surfactant concentration on the responsiveness of a thermoresponsive copolymer/surfactant mixture with potential application on "Smart" foams formulations, J. Colloid Interface Sci. 512 (2018) 455–465. doi:10.1016/j.jcis.2017.10.090.

[8]  M.J. Schick, D.J. Manning, Micelle formation in mixtures of nonionic and anionic detergents, J. Am. Oil Chem. Soc. 43 (1966) 133–136. doi:10.1007/BF02646286.

[9]  J.M. Corkill, J.F. Goodman, J.R. Tate, Micellization in mixtures of anionic and non-ionic detergents, Trans. Faraday Soc. 60 (1964) 986–995.

doi:10.1039/tf9646000986.

[10] S. Ghosh, S.P. Moulik, Interfacial and Micellization Behaviors of Binary and Ternary Mixtures of Amphiphiles (Tween-20, Brij-35, and Sodium Dodecyl Sulfate) in Aqueous Medium, J. Colloid Interface Sci. 208 (1998) 357–366. doi:10.1006/JCIS.1998.5752.

[11] Y. Moroi, N. Nishikido, M. Saito, R. Matuura, The critical micelle concentration of ionic-nonionic detergent mixtures in aqueous solutions. III, J. Colloid Interface Sci. 52 (1975) 356–363. doi:10.1016/0021-9797(75)90210-6.

[12] A. Shiloach, D. Blankschtein, Measurement and Prediction of Ionic/Nonionic Mixed Micelle Formation and Growth, Langmuir. 14 (1998) 7166–7182. doi:10.1021/la980646t.

[13] G. Sugihara, S. Nagadome, S.-W. Oh, J.-S. Ko, A review of recent studies on aqueous binary mixed surfactant systems., J. Oleo Sci. 57 (2008) 61–92. doi:10.5650/jos.57.61.

[14] J. Cross, E.J. Singer, Cationic Surfactants: Analytical and Biological Evaluation. Surfactant Science Series, Taylor & Francis, 1994.

[15] C. Carnero Ruiz, J. Aguiar, Interaction, stability, and microenvironmental properties of mixed micelles of Triton X100 and n-alkyltrimethylammonium bromides: Influence of alkyl chain length, Langmuir. 16 (2000) 7946–7953. doi:10.1021/la000154s.

[16] P.M. Holland, D.N. Rubingh, Nonideal multicomponent mixed micelle model, J. Phys. Chem. 87 (1983) 1984–1990. doi:10.1021/j100234a030.

[17] Rubingh D.N., Mixed micelle solutions, in: K. Mittal (Ed.), Solut. Chem. Surfactants, Plenum Press, Boston, MA, 1979: pp. 337–359.

[18] E.P. Schulz, G.A. Durand, Equation oriented mixed micellization modeling based on asymmetric Margules-type formulations, Comput. Chem. Eng. 87 (2016) 145–153. doi:10.1016/j.compchemeng.2015.12.026.

[19] R. Pecora, ed., Dynamic Light Scattering. Applications of Photon Correlation Spectroscopy., Plenum Press, New York and London, 1985. doi:10.1007/978-1-4613-2389-1.

[20] Malvern, User manual Zetasizer, (2013).

[21] W. Schärtl, Light Scattering from Polymer Solutions and Nanoparticle

Dispersions, Springer Berlin Heidelberg, Berlin, Heidelberg, 2007. doi:10.1007/978-3-540-71951-9.

[22]   J.H. Clint, Micellization of mixed nonionic surface active agents, J. Chem. Soc. Faraday Trans. 1 Phys. Chem. Condens. Phases. 71 (1975) 1327–1334. doi:10.1039/F19757101327.

[23]   E. Junquera, E. Aicart, Mixed micellization of dodecylethyldimethylammonium bromide and dodecyltrimethylammonium bromide in aqueous solution, Langmuir. 18 (2002) 9250–9258. doi:10.1021/la026121p.

[24]   M. Aratono, T. Takiue, Miscibility in binary mixtures of surfactants, in: M. Abe, J. Scamehorn (Eds.), Mix. Surfactant Syst., 2nd ed., Marcel Dekker, New York, 2005: pp. 13–69.

[25]   H. Hoffmann, G. Pössnecker, The Mixing Behavior of Surfactants, Langmuir. 10 (1994) 381–389. doi:10.1021/la00014a009.

[26]   I.W. Osborne-Lee, R.S. Schechter, Nonideal Mixed Micelles, in: John F. Scamehorn (Ed.), Phenom. Mix. Surfactant Syst., American Chemical Society, Washington DC, 1986: pp. 30–43. doi:10.1021/bk-1986-0311.ch002.

[27]   E.P. Schulz, J.L.M. Rodriguez, R.M. Minardi, D.B. Miraglia, P.C. Schulz, On the applicability of the regular solution theory to multicomponent systems, J. Surfactants Deterg. 16 (2013) 795–803. doi:10.1007/s11743-013-1463-3.

[28]   B. Mukhopadhyay, S. Basu, M.J. Holdaway, A discussion of Margules-type formulations for multicomponent solutions with a generalized approach, Geochim. Cosmochim. Acta. 57 (1993) 277–283. doi:10.1016/0016-7037(93)90430-5.

[29]   N.M. Van Os, J.R. Haak, L. a. M. Rupert, Physico-Chemical Properties of Selected Anionic, Cationic and Nonionic Surfactants., Elsevier, Amsterdam, 1993.

[30]   A.B. Mandal, S. Ray, A.M. Biswas, S.P. Moulik, Physicochemical studies on the characterization of Triton X 100 micelles in an aqueous environment and in the presence of additives, J. Phys. Chem. 84 (1980) 856–859. doi:10.1021/j100445a012.

[31]   L. Tianqing, G. Rong, S. Genping, Determination of the Micellar Properties of Triton X-100 by Voltammetry Method, J. Dispers. Sci. Technol. 20 (1999) 1205–1221. doi:10.1080/01932699908943845.

[32] H. Ritacco, D. Langevin, H. Diamant, D. Andelman, Dynamic surface tension of aqueous solutions of ionic surfactants: Role of electrostatics, Langmuir. 27 (2011) 1009–1014. doi:10.1021/la103039v.

[33] S.P. Moulik, M.E. Haque, P.K. Jana, A.R. Das, Micellar Properties of Cationic Surfactants in Pure and Mixed States, J. Phys. Chem. 100 (1996) 701–708. doi:10.1021/jp9506494.

[34] A.M. Mankowich, Micellar Molecular Weights of Selected Surface Active Agents, J. Phys. Chem. 58 (1954) 1027–1030. doi:10.1021/j150521a022.

[35] C.W. Dwiggins, R.J. Bolen, H.N. Dunning, ULTRACENTRIFUGAL DETERMINATION OF THE MICELLAR CHARACTER OF NON-IONIC DETERGENT SOLUTIONS 1, J. Phys. Chem. 64 (1960) 1175–1178. doi:10.1021/j100838a016.

[36] R.J. Robson, E.A. Dennis, The size, shape, and hydration of nonionic surfactant micelles. Triton X-100, J. Phys. Chem. 81 (1977) 1075–1078. doi:10.1021/j100526a010.

[37] P. Debye, Light Scattering in Solutions, J. Phys. Chem. 53 (1949) 1–8. doi:10.1021/j150466a001.

[38] J.A. Molina-Bolívar, J. Aguiar, C. Carnero Ruiz, Growth and hydration of triton X-100 micelles in monovalent alkali salts: A light scattering study, J. Phys. Chem. B. 106 (2002) 870–877. doi:10.1021/jp0119936.

[39] M. Pisárčik, F. Devínsky, E. Švajdlenka, Spherical dodecyltrimethylammonium bromide micelles in the limit region of transition to rod-like micelles. A light scattering study, Colloids Surfaces A Physicochem. Eng. Asp. 119 (1996) 115–122. doi:10.1016/S0927-7757(96)03754-5.

[40] L.A. Bulavin, V.M. Garamus, T.V. Karmazina, S.P. Shtan'ko, Micellar solutions of triton X-100: data on small-angle neutron scattering, Colloid J. Russ. Acad. Sci. 57 (1995) 902–905.

[41] P.C.Schulz, M.E.H. Vargas, J.E. Puig, Do Micelles Contribute to the Total Conductivity of Ionic Micellar Systems?, Lat. Am. Appl. Res. 25 (1995) 153–159.

[42] M.B. Sierra, P. V. Messina, M.A. Morini, J.M. Ruso, G. Prieto, P.C. Schulz, F. Sarmiento, The nature of the coacervate formed in the aqueous dodecyltrimethylammonium bromide–sodium 10-undecenoate mixtures, Colloids Surfaces A Physicochem. Eng. Asp. 277 (2006) 75–82.

doi:10.1016/J.COLSURFA.2005.11.014.

[43] M.L. Ferreira, M.B. Sierra, M.A. Morini, P.C. Schulz, Computational study of the structure and behavior of aqueous mixed system sodium unsaturated carboxylate-dodecyltrimethylammonium bromide, J. Phys. Chem. B. 110 (2006) 17600–17606. doi:10.1021/jp062599g.

[44] L.M. Kushner, W.D. Hubbard, Viscometric and Turbidimetric Measurements of Dilute Aqueous Solutions of a Non-ionic Detergent, J. Phys. Chem. 58 (1954) 1163–1167. doi:10.1021/j150522a024.

[45] A.M. Mankowich, Effect of Micellar Size on Physicochemical Properties of Surfactants, Ind. Eng. Chem. 47 (1955) 2175–2181. doi:10.1021/ie50550a042.

[46] D.G. Marangoni, A.P. Rodenhiser, J.M. Thomas, J.C.T. Kwak, Solubilization and Aggregation Numbers in Micellar Mixtures of Anionic and Cationic Surfactants with Tetraethylene Glycol and Tetraethylene Glycol Dimethyl Ether, Langmuir. 9 (1993) 438–443. doi:10.1021/la00026a013.

[47] H. Trap, Thesis, Groningen, 1953.

[48] H. Maeda, A simple thermodynamic analysis of the stability of ionic/nonionic mixed micelles, J. Colloid Interface Sci. 172 (1995) 98–105. doi:10.1006/jcis.1995.1230.

[49] K. Streletzky, G.D.J. Phillies, Temperature Dependence of Triton X-100 Micelle Size and Hydration, Langmuir. 11 (1995) 42–47. doi:10.1021/la00001a011.

[50] C. Tanford, Theory of micelle formation in aqueous solutions, J. Phys. Chem. 78 (1974) 2469–2479. doi:10.1021/j100617a012.

[51] C. Tanford, The hydrophobic effect: formation of micelles and biological membranes, Wiley, New York, 1973.

[52] H.Z. Yuan, S. Zhao, G.Z. Cheng, L. Zhang, X.J. Miao, S.Z. Mao, J.Y. Yu, L.F. Shen, Y.R. Du, Mixed Micelles of Triton X-100 and Cetyl Trimethylammonium Bromide in Aqueous Solution Studied by 1H NMR, J. Phys. Chem. B. 105 (2001) 4611–4615. doi:10.1021/jp0031303.

[53] H. Zhang, P.L. Dubin, Analysis of polydispersity of mixed micelles of TX-100/SDS and C12E8/SDS by capillary electrophoresis, J. Colloid Interface Sci. 186 (1997) 264–270. doi:10.1006/jcis.1996.4613.

[54] J.L. Rodríguez, R.M. Minardi, E.P. Schulz, O. Pieroni, P.C. Schulz, The

composition of mixed micelles formed by dodecyl trimethyl ammonium bromide and benzethonium chloride in water, J. Surfactants Deterg. 15 (2012) 147–155. doi:10.1007/s11743-011-1302-3.

[55] A. V Barzykin, M. Almgren, On the Distribution of Surfactants among Mixed Micelles, Langmuir. 12 (1996) 4672–4680. doi:10.1021/la960107t.